\documentclass[runningheads]{llncs}

\usepackage{graphicx}%
\usepackage{multirow}%
\usepackage{amsmath,amssymb,amsfonts}%
\usepackage{mathrsfs}%
\usepackage[title]{appendix}%
\usepackage{xcolor}%
\usepackage{textcomp}%
\usepackage{manyfoot}%
\usepackage{booktabs}%
\usepackage{algorithm}%
\usepackage{algorithmicx}%
\usepackage{algpseudocode}%
\usepackage{listings}%
\usepackage[T1]{fontenc}
\usepackage[utf8]{inputenc}
\usepackage{fontawesome5}
\usepackage{url}
\usepackage{hyperref} %
\usepackage{cleveref}%
\usepackage{easyReview}
\usepackage{textcomp}
\usepackage{fix-cm}

\newcommand{\jumplink}[1]{%
\href{https://wuprover.github.io/groebner_proj/docs/find/?pattern=#1\#src}{\small\faExternalLink*}%
}

\usepackage{tikz}
\usetikzlibrary{shapes, arrows.meta, positioning} 
\usepackage{wrapfig} 
\usepackage{lipsum}
\usetikzlibrary{arrows.meta, positioning, shapes.geometric, calc}
\tikzset{
    def_node/.style={
        rectangle, 
        rounded corners=3pt, 
        draw=teal!80, 
        fill=teal!10, 
        very thick, 
        minimum width=3cm, 
        minimum height=0.8cm, 
        align=center, 
        font=\sffamily\footnotesize
    },
    thm_node/.style={
        rectangle, 
        draw=blue!80, 
        fill=blue!10, 
        very thick, 
        minimum width=3cm, 
        minimum height=0.8cm, 
        align=center, 
        font=\sffamily\footnotesize
    },
    my_arrow/.style={-{Stealth[scale=1.0]}, thick, draw=gray!70}
}

\definecolor{keywordcolor}{rgb}{0.7, 0.1, 0.1}   
\definecolor{tacticcolor}{rgb}{0.0, 0.1, 0.6}    
\definecolor{commentcolor}{rgb}{0.4, 0.4, 0.4}   
\definecolor{symbolcolor}{rgb}{0.0, 0.1, 0.6}    
\definecolor{sortcolor}{rgb}{0.1, 0.5, 0.1}      

\lstnewenvironment{leancode}{\lstset{frame=single,language=lean,abovecaptionskip=-\medskipamount,basicstyle={\ttfamily\footnotesize}}}{}

\newcommand{\lean}{\lstinline}
\usepackage[T1]{fontenc}
\usepackage{graphicx}
\BeforeBeginEnvironment{leancode}{\smallskip}

\usepackage{diagbox}

\usepackage{xifthen}
\newcommand*{\leandef}[2][]{\href{https://wuprover.github.io/MonomialOrderedPolynomial/docs/find/?pattern=#2\#doc}{\ifthenelse{\isempty{#1}}{\lean{#2}}{#1}}}

\newcommand\leantodo[1]{}
\lstdefinelanguage{json}{
    basicstyle=\ttfamily\scriptsize, 
    numbers=none,                    
    showstringspaces=false,
    breaklines=true,
    frame=single,                    
    rulecolor=\color{gray!30},       
    backgroundcolor=\color{gray!2},  
    stringstyle=\color{blue},
    keywordstyle=\color{black},      
}
\begin{document}
\title{Automated Tactics for Polynomial Reasoning in Lean~4} 
%
%







\author{Hao Shen\inst{1} \and
Junyu Guo\inst{2} \and
Junqi Liu\inst{1} \and
Lihong Zhi \inst{1}
\authorrunning{F. Author et al.}
%
\institute{State Key Laboratory of Mathematical Sciences, Academy of Mathematics and Systems Science, University of Chinese Academy of Science, Beijing, 100190, China \\
 \email{\{shenhao24,liujunqi,lzhi\}@amss.ac.cn,}
   \and
Institute of Logic and Cognition, Department of Philosophy, Sun Yat-sen University, 510275, Guangdong, China \\
\email{guojy228@mail2.sysu.edu.cn}}}
\maketitle              
\begin{abstract}
Applying Gröbner basis theory to concrete problems in Lean~4 remains difficult since the current formalization of multivariate polynomials is based on a non-computable representation and is therefore not suitable for efficient symbolic computation. As a result, computing Gröbner bases directly inside Lean is impractical for realistic examples.
To address this issue, we propose a certificate-based approach
 that combines external computer algebra systems, such as SageMath or SymPy, with formal verification in Lean~4. Our approach uses a computable representation of multivariate polynomials in Lean to import and verify externally generated Gröbner basis computations. The external solver carries out the main algebraic computations, while the returned results are verified inside Lean.
Based on this method, we develop automated tactics that transfer polynomial data between Lean and the external system and certify the returned results. These tactics support tasks such as remainder verification, Gröbner basis checking, ideal equality, and ideal or radical membership. This work provides a practical way to integrate external symbolic computation into Lean~4 while preserving the reliability of formal proof.
\end{abstract}

\section{Introduction}

Gröbner basis, first introduced by Buchberger~\cite{BUCHBERGER2006475}, is a cornerstone of computational commutative algebra and algebraic geometry. It provides a unified algorithmic framework for classical problems, such as deciding  ideal membership  and solving systems of polynomial equations. 

Lean 4 is an open-source interactive theorem prover based on dependent type theory, featuring a small trusted kernel \cite{moura2021lean}. Its rapid growth is epitomized by Mathlib \cite{mathlib}, a unified and extensive mathematical library built by a collaborative community of mathematicians and computer scientists. In the Lean 4 ecosystem, Gröbner basis theory has been formalized by Guo et al.~\cite{guo2026formalizinggrobnerbasistheory}. However, the formalization relies on the non-computable structure \lean{MvPolynomial}, which results that applying Gröbner basis theory to concrete computational tasks remains a significant challenge in Lean. 

Instead of computing a Gröbner basis in Lean, we assign the heavy computation to an external computer algebra system (CAS). Our approach is inspired by a series of studies dedicated to combining the rigor of interactive theorem provers with the computational efficiency of CAS. The concept of delegating algebraic computations to external tools has proven successful across various proof assistants. For instance, Delahaye and Mayero pioneered this methodology in Coq~\cite{coq} (renamed to Rocq in recent years) by interfacing with Maple~\cite{char2013maple} to handle field operations and simplify algebraic expressions, demonstrating that external solvers can significantly streamline formal proofs without compromising rigor~\cite{delahaye2005dealing}. Similarly, Seddiki et al. extended the capabilities of HOL Light~\cite{harrison2009hol} by integrating it with symbolic and numerical computation engines, further validating the approach where the interactive theorem prover certifies results provided by an untrusted CAS~\cite{seddiki2015enabling}. Besides, Lewis and Wu~\cite{lewis2022bi} established a bidirectional extensible interface between Lean~3 and Mathematica~\cite{wolfram2003mathematica}. 


We use SageMath~\cite{sage2020sagemath} or SymPy~\cite{meurer2017sympy} as the external solver, taking advantage of their open-source infrastructures and powerful symbolic computation capabilities. However, computing a result externally is only the first step. The result must still be rigorously verified within Lean~4. Direct verification via \leandef{MvPolynomial} is computationally impractical since its highly generic design supports abstract mathematical development rather than efficient large-scale symbolic computation. To address this bottleneck, we use a computable representation of a polynomial developed by Guo et.al.~\cite{wuprover_mop_2026}. Finally, we encapsulate this entire workflow into tactics, allowing users to solve algebraic goals automatically without managing the underlying complexity. The source code is publicly available at \url{https://github.com/WuProver/GroebnerTactic}.


\section{Lean-CAS Interface}
\label{expr_and_term}
Lean~4 is based on the Calculus of Inductive Constructions (CIC), a dependent type theory in which mathematical objects, types, and proofs are represented uniformly as \textit{terms}. In this setting, proof verification is just type checking, and Lean further supports proof development through tactic-based automation.

While users typically work with Lean through its term language and tactic language, the implementation of such automation tactics takes place at the metaprogramming level. In our development, we implement tactics directly at this level. More precisely, our tactics operate on Lean's internal expression type \leandef[\lean{Expr}]{Lean.Expr}, which represents logical expressions as abstract syntax trees encoding CIC terms. This inductive type includes constructors for the main syntactic forms of the language, such as free and bound variables (\texttt{fvar}, \texttt{bvar}), constants (\texttt{const}), function application (\texttt{app}), binders including lambda abstractions and dependent function types (\texttt{lam}, \texttt{forallE}), universe expressions, and metavariables (\texttt{mvar}) representing open proof goals.

Our tactics inspect, manipulate, and construct these \lean{Expr} objects to implement proof automation. The constructors and operations used in our implementation will be introduced as needed in the following sections. A central component of this implement is the interface between Lean~4 and the external computer algebra system, which we will describe in the remainder of this section.

\subsection{Lean to CAS}

The first stage of the pipeline extracts the relevant algebraic data from Lean's internal expression representation, \lean{Expr}, and serializes it into a form that can be processed by the external computer algebra system.

In Lean~4, terms are represented internally as abstract syntax trees. To parse and construct these trees reliably, we use the \texttt{quote4} library~\cite{quote4} to match and extract Lean expressions against typed quotation patterns from standard algebraic objects and operations, such as natural number literals and ring operations. This provides a type-safe way to inspect the syntax of terms and ensures that the extracted data is well typed by construction.

Once the polynomials have been extracted, they are serialized and embedded into a script that can be executed by an external solver.

\subsection{CAS to Lean}
\label{sage_to_lean}
The computations are carried out by external computer algebra systems, and the results are returned in JSON format. To support efficient communication between the external solver and Lean, we adopt a sparse representation of multivariate polynomials with rational coefficients. As shown in \Cref{poly_json}, each polynomial is serialized as a list of monomials.
\begin{lstlisting}[language=json, caption={Representation of $f = \frac{3}{4}x_1^2x_2 - 7x_3^5 + 2$}, label={poly_json}]
[
  {"c": [3, 4],  "e": [[1, 2], [2, 1]]},
  {"c": [-7, 1], "e": [[3, 5]]},
  {"c": [2, 1],  "e": []}
]
\end{lstlisting}

On the Lean side, we introduce lightweight, serializable representations for coefficients, variables, monomials, and polynomials, together with reification functions that turn each of them into a Lean \lean{Expr}. It encodes rational coefficients as integer–natural pairs, variables as index–exponent pairs, monomials as a coefficient bundled with an array of such pairs, and polynomials as lists of monomials. The reification functions build up the corresponding \lean{MvPolynomial σ R} expression compositionally: variables become powers of indeterminates, monomials are assembled by multiplying a constant term with its variable factors, and polynomials are obtained by summing their reified monomials.
Using these structures and conversion functions, we translate the JSON output produced by an external solver into an \lean{Expr} object in Lean, and then apply \leandef{Lean.PrettyPrinter.delab} to convert it into a \leandef[\lean{Syntax}]{Lean.Syntax} object that can serve as a certificate for other tactics.

\section{Three Tactics for Gröbner Basis Reasoning}


In this section, we introduce three tactics for automated algebraic reasoning in Lean~4. These tactics complement existing automation, such as \lean{grind}, by providing dedicated support for algebraic reasoning tasks. At present, our implementation supports only the lexicographic monomial order and is restricted to problems formulated over the polynomial ring \(\mathbb{Q}[x_0, \ldots, x_n]\).




\subsection{Proving Ideal Equality: \lean{idealeq} }
\label{idealeq}

For some kinds of verifications requiring a Gröbner basis, it is essential to verify in Lean that the externally computed  Gröbner basis $G$ of $\langle B \rangle$ is a Gröbner basis, but it does not by itself imply that $G$ generates the same ideal as the original generator $B$. Formal algebraic reasoning requires us to prove the extensional identity $\langle G \rangle = \langle B \rangle$. To automate this task, we implement the tactic \lean{idealeq}, which proves the equality of two finitely generated ideals by reducing it to ideal membership certificates.

Suppose $
I = \langle f_1,\dots,f_n\rangle,
J = \langle g_1,\dots,g_m\rangle.
$
To prove $I = J$, it is enough to establish the two inclusions $I \subseteq J$ and $J \subseteq I$. Taking proving $I \subseteq J$ as an example, it suffices to show that each generator $f_i$ of $I$ lies in $J$. This membership condition is equivalent to exhibiting polynomials
$
c_{i1},\dots,c_{im} \in \mathbb{Q}[x_0,\ldots,x_n]
$
such that $
f_i = \sum_{j=1}^m c_{ij} g_j.
$
Thus, the problem of ideal inclusion is reduced to finding explicit coefficient polynomials witnessing that every generator of one ideal lies in the span of the generators of the other.

In our implementation, these coefficient polynomials are computed by an external solver and returned to Lean as certificates. On the Lean side, we implement a tactic \lean{submodule_span} that takes such coefficients as input and reduces the membership goal to an equality check,  which is a concrete polynomial identity testing (PIT) problem and can be solved by the computable representation. The tactic \lean{idealeq} establishes $I = J$ by proving both inclusions that every $f_i \in J$ and every $g_j \in I$.

\begin{example}\label{ideal_eq_example}
Consider the lexicographic order $x_0 > x_1$ in the polynomial ring $\mathbb{Q}[x_0, x_1]$. Let $I = \langle x_0 + x_1^2, x_1^2 \rangle$ and $J = \langle x_0, x_1^2 \rangle $, then $I = J$.
\end{example}

The \lean{idealeq} tactic can automatically solve the ideal equality problem in \Cref{ideal_eq_example} as follows. 
\begin{leancode}
example : Ideal.span ({X 0 + X 1^ 2, X 1 ^ 2}) =
      Ideal.span ({X 0, X 1 ^ 2} : Set (MvPolynomial (Fin 2) ℚ)) := by
  idealeq
\end{leancode}


\subsection{Solving Problems Involving Gröbner Bases: \lean{gb_solve}}

We consolidate the algebraic reasoning capabilities regarding Gröbner Bases into a single unified tactic, \lean{gb_solve}, which automates four classes of goals. 
The tactic first inspects the current proof goal in Lean's metaprogramming environment, performs pattern matching on its structure, and then dispatches it to the appropriate backend procedure.
Concretely, \lean{gb_solve} recognizes four forms of goals, which we describe below. The notions of polynomial remainder, Gröbner basis, 
$S$-polynomial, and Buchberger's criterion have been formalized in Lean by Guo et al.~\cite{cox1997ideals,guo2026formalizinggrobnerbasistheory}.


\subsubsection{Certification of Polynomial Remainders}
When the goal has the form \lean{lex.IsRemainder f B r}, the tactic delegates the computation to an external solver, which returns quotient polynomials $q_i$ satisfying
 \[f - r = \sum_i q_i \cdot b_i\]
 back into Lean as introduced in section \ref{expr_and_term}. 
At this point, the original goal is reduced to verifying a comparison between the degrees of polynomials, together with a polynomial identity, both of which the tactic can solve automatically.


\begin{example}
\label{remainder_example}

Consider the lexicographic order $x_0 > x_1$ on the polynomial ring $\mathbb{Q}[x_0,x_1]$. Let $
f = x_0^2x_1 + x_0x_1^2, 
G = \{x_0x_1 - 1\}.
$
Then the remainder of $f$ upon division by $G$ is $x_0 + x_1$.
\end{example}

The tactic \lean{gb_solve} automatically proves in \Cref{remainder_example} that \(r\) is the remainder of \(f\) divided by \(G\).

\begin{leancode}
example : lex.IsRemainder
    (X 0 ^ 2 * X 1 + X 0 * X 1 ^ 2 : MvPolynomial (Fin 3) ℚ)
    {X 0 * X 1 - 1} (X 0 + X 1) := by gb_solve
\end{leancode}

\subsubsection{Gröbner Basis Verification}
When the goal has the form \lean{lex.IsGroebnerBasis G I}, the verification proceeds in two steps. 
The first step uses Buchberger's criterion~\cite{cox1997ideals}: to prove that \(G\) is a Gr\"obner basis of the ideal it generates, it suffices to show that every \(S\)-polynomial of pairs \(g_i, g_j \in G\) reduces to zero with respect to \(G\). In our implementation, this is checked automatically using the remainder certification procedure above.

This, however, proves only that $G$ is a Gröbner basis of the ideal $\langle G\rangle$. To show that $G$ is a Gröbner basis for the intended target ideal $I$, we must further establish that $\langle G\rangle = I.$
We verify this equality using the tactic \lean{idealeq}. Once Buchberger's criterion and the ideal equality have both been certified, the tactic combines them to conclude the goal.

\begin{example}
\label{gb_example}
Consider the lexicographic order $x_0 > x_1$ in the polynomial ring $\mathbb{Q}[x_0, x_1]$. Let $I = \langle x_0 + x_1,\, x_0 x_1 - 1 \rangle$. Then $G = \{x_0 + x_1,\, x_1^2 + 1\}$ is a Gröbner basis of $I$.
\end{example}
\begin{leancode}
example : lex.IsGroebnerBasis
    ({X 0 + X 1, X 1 ^ 2 + 1} : Set (MvPolynomial (Fin 2) ℚ))
    (Ideal.span {X 0 + X 1, X 0 * X 1 - 1}) := by gb_solve
\end{leancode}

\subsubsection{Ideal Membership Problem}
\label{ideal_mem_tac}

A fundamental property of Gröbner bases is that they provide an effective method for solving the ideal membership problem. For a goal of the form \lean{f ∈ Ideal.span B}, \lean{gb_solve} invokes an external computation tool to obtain an ideal-membership certificate, namely polynomials $c_i$ such that $
f = \sum_i c_i \cdot s_i.$
These coefficients are then deserialized into Lean expressions and passed to our tactic \lean{submodule_span}, which reduces the original goal to an explicit polynomial identity. This identity is then verified inside Lean's kernel.

\begin{example}
\label{ideal_mem_pos_example}
Consider the lexicographic order $x_0 > x_1$ in the polynomial ring $\mathbb{Q}[x_0, x_1]$, then $1 \in \langle x_0,\ 1 - x_1 x_0 \rangle$.
\end{example}
\begin{leancode}
example : 1 ∈ Ideal.span ({X 0, 1 - X 1 * X 0} : Set <| MvPolynomial (Fin 2) ℚ) := by gb_solve
\end{leancode}


For a goal of the form \lean{¬(f ∈ Ideal.span B)}, the tactic proceeds differently. It first computes a Gröbner basis $G$ of the ideal $\langle B \rangle$, and then computes the remainder $r$ of $f$ upon division by $G$. 
The membership condition $f \in \langle B \rangle$ holds if and only if $r = 0$ \cite{BUCHBERGER2006475}. Therefore, to prove the negated goal, it suffices to certify that $r \neq 0$. In this way, the non-membership problem is again reduced to a concrete polynomial computation, and ultimately to comparing polynomial degrees along with a polynomial identity testing (PIT) problem over the computable polynomial representation.

\begin{example}
\label{ideal_mem_neg_example}

Consider the polynomial ring $\mathbb{Q}[x_0, x_1]$ with lexicographic order $x_0 > x_1$. Let
$
f = x_0 + x_1, 
I = \langle x_0 + x_1^2,\; x_1^2 \rangle.
$
Then $f$ does not belong to $I$.
\end{example}
\begin{leancode}
example : X 0 + X 1 ∉ Ideal.span ({X 0 + X 1^ 2, X 1 ^ 2} : Set (MvPolynomial (Fin 2) ℚ)) := by gb_solve
\end{leancode}

\subsubsection{Radical Membership Problem}



For a goal of the form \lean{f ∈ (Ideal.span S).radical}, the tactic \lean{gb_solve} first applies \lean{Ideal.mem_radical_iff} to reduce the problem to proving that $f^n \in \langle S \rangle$ for some exponent $n$. It then queries an external solver for such a witness exponent together with the corresponding ideal membership certificate. Once these data are returned, the tactic reduces the goal to an ordinary ideal membership problem. 

\begin{example}
\label{radical_pos_example}
Consider the lexicographic order $x_0 > x_1$ in the polynomial ring $\mathbb{Q}[x_0, x_1]$. Let $f = (x_0 + x_1)(x_0-x_1)$ and $I = \langle x_0^2, x_1^2\rangle$, then $f \in \sqrt{I}$.
\end{example}
\begin{leancode}
example : (X 0 - X 1) * (X 0 + X 1) ∈ (Ideal.span ({X 0^2, X 1^2} : Set (MvPolynomial (Fin 2) ℚ))).radical := by gb_solve
\end{leancode}

For the negated goal, the tactic applies Rabinowitsch's method~\cite{cox1997ideals}, which we have formalized in Lean~4. More precisely, it constructs the extended ideal $
\langle B,\; 1 - t f \rangle \subseteq k[x_i,t]$
by lifting the generators in $B$ to the larger polynomial ring with an extra variable $t$ and adjoining the auxiliary polynomial \(1-tf\). By Rabinowitsch's method, proving that $
1 \notin \langle B,\; 1-tf \rangle
$
establishes that \(f \notin \sqrt{\langle B\rangle}\). The resulting goal is therefore reduced to the non-membership of \(1\) in an ideal, which is handled by the procedure described above.
\begin{example}
\label{radical_neg_example}
Consider the polynomial ring $\mathbb{Q}[x_0, x_1]$ equipped with the lexicographic order $x_0 > x_1$. Let $f = x_0$ and $I = \langle x_0 + x_1 \rangle.$ Then \(f \notin \sqrt{I}\).
\end{example}

\begin{leancode}
example : X 0 ∉ (Ideal.span ({X 0 + X 1} : Set (MvPolynomial (Fin 2) ℚ))).radical := by gb_solve
\end{leancode}

%

\subsection{Computing a Gröbner Basis and Adding It as a Hypothesis: \lean{add_gb_hyp}}


The procedures described above assume that a Gröbner basis \(G\) has already been provided. In practice, however, one usually starts with an arbitrary generating set \(B\) and must first compute a Gröbner basis for the ideal it generates. To support this workflow, we implement the tactic \lean{add_gb_hyp}. Given a generating set \(B\), this tactic calls an external solver to compute a Gröbner basis \(G\) of the ideal \(\langle B \rangle\).

Once the candidate basis \(G\) is returned, \lean{add_gb_hyp} certifies in Lean that
\lean{lex.IsGroebnerBasis G} (\lean{Ideal.span B})
holds, using the verification procedure described in the previous subsection. In this way, the tactic does not merely import the output of the external computation but turns it into a formally certified result.

After the certification, the tactic inserts the proof of
\lean{lex.IsGroebnerBasis G} (\lean{Ideal.span B})
into the local proof context as a named hypothesis. This hypothesis can then be reused in subsequent proof steps, allowing later tactics to work directly with the computed Gröbner basis.

For example, in \Cref{gb_example}, if the Gröbner basis of \(I\) is not known in advance, we may first invoke \lean{add_gb_hyp} to compute and certify such a basis before proceeding with the rest of the proof.
\begin{leancode}
example : ∃ G : Set (MvPolynomial (Fin 2) ℚ), lex.IsGroebnerBasis G
  (Ideal.span ({X 0 + X 1, X 0 * X 1 - 1} : Set <| MvPolynomial (Fin 2) ℚ)) := by
  add_gb_hyp h ({X 0 + X 1, X 0 * X 1 - 1} : Set (MvPolynomial (Fin 2) ℚ))
  exact Exists.intro _ h
\end{leancode}


\subsection{Supported Computation Modes}

To support calls from Lean~4 tactics to external computer algebra systems, we implement a unified runner framework with three backends: local SageMath, SageMath accessed through an API, and local SymPy. The backend to be used is controlled by the Lean~4 option \lean{gb_tactic.backend}. By default, the system uses local SageMath.

The interface between the tactic layer and the computation backends is given by a common task descriptor type \lean{GbTask}. Its constructors represent the various algebraic operations required in our workflow, including polynomial reduction, Gröbner basis computation, reduction to  normal form, ideal membership testing, and radical membership testing.

Each backend implements the same execution pattern. A task represented in type \lean{GbTask} is first mapped into a backend-specific script name along with the required command-line arguments. The task is then executed by spawning the corresponding external process through \leandef{IO.Process.spawn}. Finally, the backend returns the output, which will subsequently be parsed and used by the tactic in Lean.

Taking \Cref{ideal_eq_example} as an example, explicitly setting \lean{gb_tactic.backend} to \texttt{1} directs the computation through the SageMath API, which is particularly convenient in environments where SageMath is not installed locally.
\begin{leancode}
set_option gb_tactic.backend 1 in
example : Ideal.span ({X 0 + X 1^ 2, X 1 ^ 2}) =
    Ideal.span ({X 0, X 1 ^ 2} : Set (MvPolynomial (Fin 2) ℚ)) := by
    idealeq
\end{leancode}
Setting it to \texttt{2} delegates to a SymPy installed locally instead. 
\begin{leancode}
set_option gb_tactic.backend 2 in
example : Ideal.span ({X 0 + X 1^ 2, X 1 ^ 2}) =
    Ideal.span ({X 0, X 1 ^ 2} : Set (MvPolynomial (Fin 2) ℚ)) := by
    idealeq
\end{leancode}

\section{Conclusion}

In this paper, we presented a framework for automated algebraic reasoning in Lean~4 through integration with external computer algebra systems. At the metaprogramming level, we developed an interface between Lean and external solvers, enabling the extraction, serialization, and certification of algebraic data and computational results. Building on this infrastructure, we introduced three automated tactics: \lean{idealeq}, \lean{gb_solve}, and \lean{add_gb_hyp} that support a range of algebraic reasoning tasks, including ideal equality, remainder certification, Gröbner basis verification, ideal membership problems, and radical ideal membership problems.

A promising direction for future work is to implement the relevant symbolic algebraic computations directly and verifiably within Lean, thereby removing the dependence on external solvers and further strengthening the reliability of the overall framework.


\begin{credits}
\subsubsection{\ackname}This work was supported by the Strategic Priority Research Program of Chinese Academy of Sciences under Grant XDA0480502  and the National Key R\&D Program of China 2023YFA1009401. Junyu Guo is encouraged and supported by his supervisor, Xishun Zhao, to work on this project.
\end{credits}

\bibliographystyle{splncs04}
\bibliography{references}
%






\end{document}